\documentclass[%
 superscriptaddress,
 longbibliography,
 amsmath,amssymb,
 reprint,%
]{revtex4-1}

\usepackage{graphicx}
\usepackage{dcolumn}
\usepackage{bm}
\usepackage{gensymb}
\usepackage[dvipsnames]{xcolor}
\usepackage[colorlinks=true,citecolor=blue]{hyperref}

\usepackage[utf8]{inputenc}
\usepackage[T1]{fontenc}
\usepackage{mathptmx}
\usepackage{etoolbox}

\graphicspath{{figures/}}

\makeatletter
\def\@email#1#2{%
 \endgroup
 \patchcmd{\titleblock@produce}
  {\frontmatter@RRAPformat}
  {\frontmatter@RRAPformat{\produce@RRAP{*#1\href{mailto:#2}{#2}}}\frontmatter@RRAPformat}
  {}{}
}%
\makeatother

\begin{document}

\preprint{AIP/123-QED}


\title{Higher-order ferromagnetic resonances in periodic arrays of synthetic-antiferromagnet nanodiscs}

\author{V.~Yu.~Borynskyi}
\affiliation{Institute of Magnetism of NAS of Ukraine and MES of Ukraine, 36-b Akad. Vernadsky blvd., Kyiv 03142, Ukraine}

\author{D.~M.~Polishchuk}
\affiliation{Institute of Magnetism of NAS of Ukraine and MES of Ukraine, 36-b Akad. Vernadsky blvd., Kyiv 03142, Ukraine}
\affiliation{Nanostructure Physics, Royal Institute of Technology, Stockholm 10691, Sweden}

\author{A.~K.~Melnyk}
\affiliation{Institute for Sorption and Problems of Endoecology of NAS of Ukraine, 13 Naumov str., Kyiv 03164, Ukraine}

\author{A.~F.~Kravets}
\affiliation{Institute of Magnetism of NAS of Ukraine and MES of Ukraine, 36-b Akad. Vernadsky blvd., Kyiv 03142, Ukraine}
\affiliation{Nanostructure Physics, Royal Institute of Technology, Stockholm 10691, Sweden}

\author{A.~I.~Tovstolytkin}
\affiliation{Institute of Magnetism of NAS of Ukraine and MES of Ukraine, 36-b Akad. Vernadsky blvd., Kyiv 03142, Ukraine}

\author{V.~Korenivski}
\affiliation{Nanostructure Physics, Royal Institute of Technology, Stockholm 10691, Sweden}



\begin{abstract}
We investigate spin dynamics in nanodisc arrays of synthetic-antiferromagnets (SAF) made of Py/NiCu/Py trilayers, where the NiCu spacer undergoes a Curie transition at about 200~K. The observed ferromagnetic resonance spectra have three distinct resonance modes at room temperature, which are fully recreated in our micromagnetic simulations showing also how the intra-SAF asymmetry can be used to create and control the higher-order resonances in the structure. Below the Curie temperature of the spacer, the system effectively transitions into a single-layer nanodisc array with only two resonance modes. Our results show how multi-layering of nano-arrays can add tunable GHz functionality relevant for such rapidly developing fields as magnetic meta-materials, magnonic crystals, arrays of spin-torque oscillators and neuromorphic junctions.
\end{abstract}

\maketitle


The last decade has seen large research efforts focused on collective spin dynamics of magnonic crystals and logic circuits aimed for various high-speed applications~\cite{Chumak2015,Liu2013,Lenk2011,Haldar2016}. Controllable size of the individual elements used in such metamaterials, down to the 10~nm range, enables rich higher-order FMR dynamics such as standing spin waves and edge-mode resonances. The typical geometry of a periodic array offers high flexibility in engineering novel properties. At the same time, the increased complexity requires systematic investigations of various effects and parameters involved~\cite{Saha2012,Liu2008,Schneider2007,Verba2020}. 

Relatively simple periodic arrays of single-layer ferromagnetic nanodiscs exhibit anisotropic spin wave modes (commonly, with four-fold in-plane anisotropy) affected by the inter-element dipolar interactions~\cite{Mathieu1997,Kakazei2006,Carlotti2019}. The properties become more intriguing for arrays of multilayered elements such as SAF's, which are of interest in antiferromagnetic spintronics~\cite{Duine2018} as well as magnonics~\cite{Etesamirad2021}. Their collective spin dynamics is dictated by the inter-element interactions, the intra-particle dipolar coupling, and the interplay of the two~\cite{Pauselli2017,Kamimaki2020,Chatterjee2018}.

Arrays of multilayered nanomagnets can exhibit distinct spin dynamic modes not found in single-layer systems. The intra-particle dipolar interactions lead to complex demagnetizing field distributions, altering the collective spin dynamics in such systems, resulting in mode-frequency shifts~\cite{Talapatra2021} or even splitting of the resonant modes~\cite{Carlotti2015}. For example, a systematic micromagnetic study performed by Pauselli~\textit{et al.}~\cite{Pauselli2017} for a sub-40-nm magnetic tunnel junction read head explained additional higher-order modes as originating from a strong localization of spin excitations to the junction's edges, where an excitation in one of the junction's magnetic layers drives a localized resonance in the neighboring, dipole-coupled magnetic layer. Multi-layering is thus a capable tool for varying spin wave modes and therefore spectral modification~\cite{Haldar2020,Begari2018}, additional to element grouping or sequencing in arrays of single-layer nanoparticles~\cite{Carlotti2015,Gartside2021}. Furthermore, particle shape imperfections unavoidable on the experiment, such as edge deviations~\cite{Nembach2021} or a sidewall angle~\cite{Maranville2007}, can have a significant effect on the spin dynamics of the system. Strongly localized edge-mode excitations are highly sensitive to this kind of distortions, which can lead to lifting of their degeneracy for both single-layer~\cite{Chia2012} and multilayered~\cite{Zhang2019} systems. While arrays of singe-layer nanoparticles are well studied~\cite{Hu2015,Ding2012,Shaw2009}, a systematic investigation of arrays of multilayered nano-elements is still pending.

In this work, we investigate higher-order FMR excitations in periodic arrays of multilayered SAF nanodiscs, designed with an in-situ thermal control of the interlayer coupling capable of making the nanodiscs effectively single-layered. This in-situ control of the SAF versus single-layer character of the array elements allows us to identify the SAF-unique properties from a direct comparison of the FMR spectra measured above and below the Curie point of the spacer. Using detailed micromagnetic simulations, we explain the occurrence, positions, and anisotropy of the measured resonances and how they are affected by strong localizations of the spin excitations and a complex interplay between the inter- and intra-particle interactions, pronounced in SAF-based nanostructures. 

We focus our investigation on arrays of SAF nanodiscs Py(7~nm)/NiCu(10~nm)/Py(7~nm) with periodicity 250~nm and nominal diameter $d~=~150$~nm. Here, Py and NiCu stand for Ni$_{80}$Fe$_{20}$ (Permalloy) and Ni$_{60}$Cu$_{40}$, respectively. The spacer is paramagnetic at room temperature (RT) and mediates no direct exchange between the outer Py layers, which interact through magnetostatic coupling only, thereby forming a SAF element. At low temperature, the spacer is ferromagnetic, which allows in-situ switching between the SAF and the single-layer states in the system~\cite{Kravets2012,Kravets2014}. Multilayers Py/NiCu/Py were grown by magnetron sputtering (Orion, AJA Int.) onto thermally oxidized Si (001) wafers at room temperature. The multilayers were then used to fabricate arrays of SAF nanodiscs using electron-beam lithography (Voyager, Raith Inc.). The arrays were etched using Ar-plasma via a double-layer hard mask, TaN(60~nm)/Al(20~nm). Here, thin Al was used as a lift-off layer to form a mask (with the nominal diameter $d_0~=~150$~nm) for selective reactive plasma etching with SF6 of the thick TaN layer underneath. Additional array geometries were fabricated and tested for calibration and verification purposes, in the process of optimizing the SAF arrays, with most interesting properties presented herein. 

\begin{figure}
\includegraphics[width=8 cm]{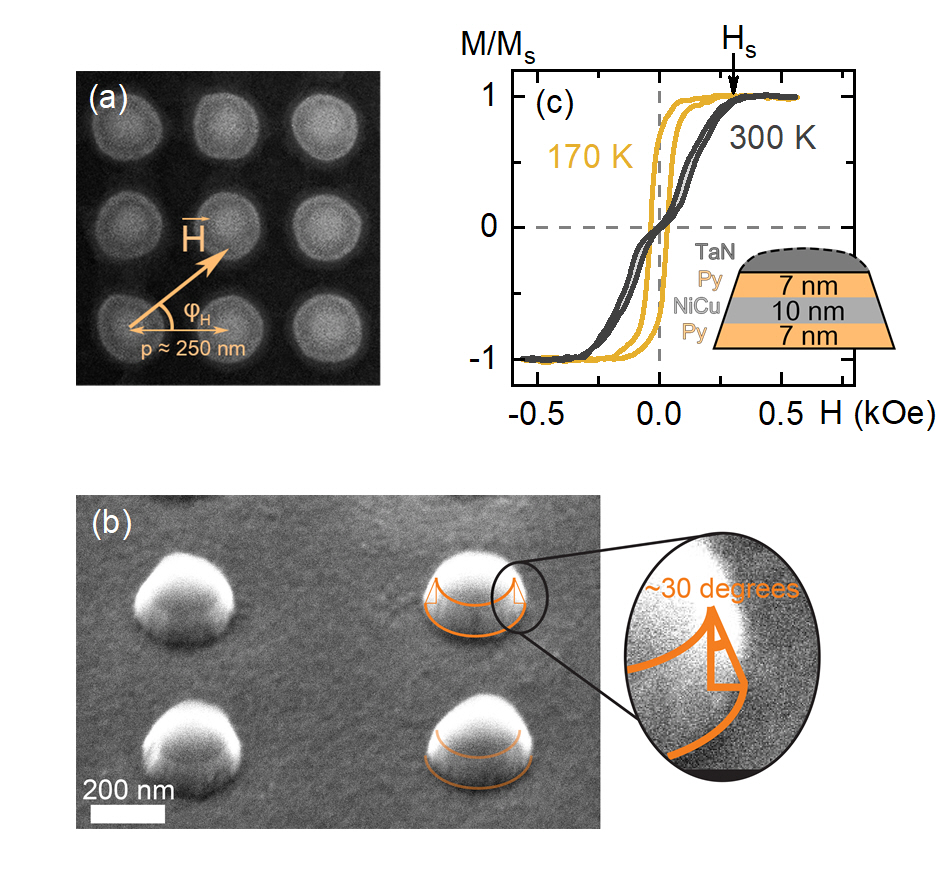}
\caption{(a) Top-view SEM image of select region in planar array of 150-nm SAF nanodiscs (Py/NiCu/Py). Inset indicates particle periodicity and in-plane orientation of external magnetic field. (b) 45$\degree$ SEM view of same sample. (c) Typical M-H loops, measured using MOKE at 300 and 170~K; inset illustrates vertical cross-section of single SAF nanodisc. TaN is used as etching mask and for capping.}
\label{fig1}
\end{figure}

SEM images of the fabricated arrays in Fig.~\ref{fig1}(a--b) show good uniformity in the element size (the spread is less than 5~\%) and good edge morphology. Angled SEM in Fig.~\ref{fig1}(b) shows a side-wall angle of 20--30$\degree$, formed in the process of plasma etching, making the effective diameter of the top Py disc about 15~\% smaller than that of the bottom disc. The top dome-like cap, visible as the central concentric circles in the top-view and contrasted-top in the side-view SEM images, Fig.~\ref{fig1}(a--b), is the remaining TaN mask that protects the fabricated particles from oxidization.

Cavity-FMR measurements were carried out using an X-band ELEXSYS E500 spectrometer (Bruker Inc.) equipped with an automatic goniometer for precise angle control. The measurements were performed at a constant operating frequency of 9.36~GHz. The magnetic volume of the arrays of area $1.4\times1.4$~mm$^2$ was large enough for a high signal-to-noise ratio. Additionally, magnetostatic properties of the arrays were measured using a custom-made variable-temperature magneto-optical Kerr-effect (MOKE) setup.

Figure~\ref{fig1}(c) shows two characteristic MOKE loops measured in-plane ($\varphi_H = 0\degree$). Zero remnant magnetization at $H = 0$ and relatively high saturation field $H_s \approx 300$~Oe for RT loop indicate a pronounced antiferromagnetic-type interlayer coupling of the top and bottom magnetic moments of the individual SAF nanodiscs. The double-step shape and small hysteresis can arise from the presence of magnetic anisotropy~\cite{Polishchuk2018} as well as a non-ideal circular shape of the nanodiscs~\cite{Koop2017,Holmgren2017}.

\begin{figure*}
\includegraphics[width=17 cm]{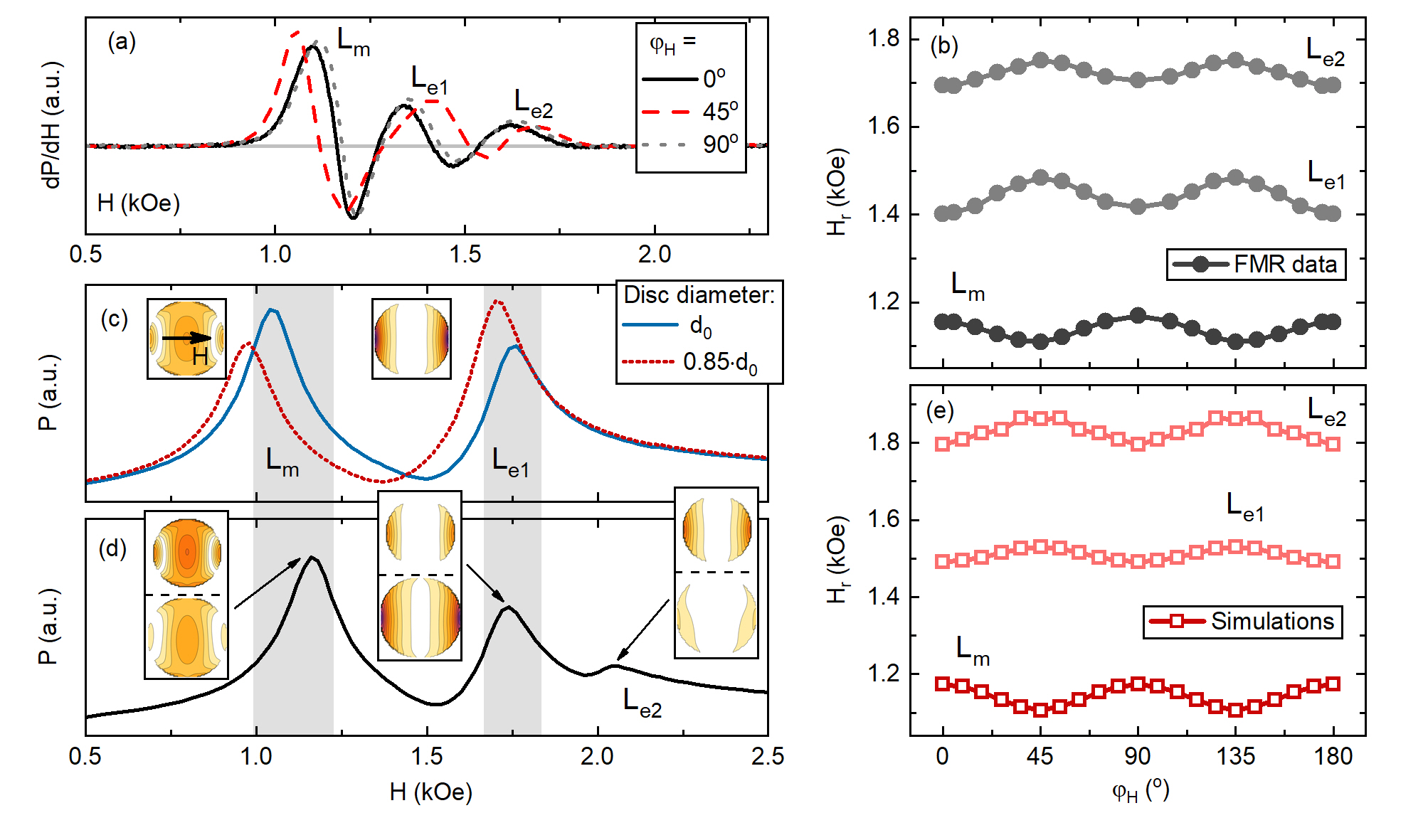}
\caption{(a) FMR spectra measured at RT for 150-nm SAF array at 0, 45, and 90$\degree$-oriented external field. L$_m$, L$_{e1}$ and L$_{e2}$ mark resonance lines attributed to main and two edge-modes, respectively. (b) Resonance fields of individual modes versus in-plane angle. (c) FMR spectra calculated for individual single-layer discs with $d_0 = 150$~nm and $d \approx 0.85d_0$, corresponding to bottom and top layers of SAF nanodiscs. Insets show excitation regions for two eigen-modes. (d) Calculated FMR spectrum for individual SAF particle with $d_0$ and $d \approx 0.85d_0$. Insets display excitation regions in both layers of SAF nanodisc. All spectra were calculated with in-plane field at $\varphi_H = 0\degree$ (black arrow in left inset to c). (e) In-plane angular dependence of resonance fields simulated for SAF disc array with $d_0 = 150$~nm, $d \approx 0.85d_0$, and array periodicity 200~nm.}
\label{fig2}
\end{figure*}

Figure~\ref{fig2}(a) shows the FMR spectra measured at RT for the array of 150-nm SAF nanodiscs with different in-plane orientations of the external magnetic field, $\varphi_H = 0, 45$, and 90$\degree$. Each curve shows three well-defined resonance lines marked as L$_m$, L$_{e1}$, and L$_{e2}$. We associate the L$_m$ line with the quasi-uniform (hereafter \textit{main}) FMR mode since its position coincides with that of the resonance line for a continuous Py film. With respect to the position of L$_m$, lines L$_{e1}$ and L$_{e2}$ are observed at higher resonance fields. A similar higher-field mode, along with the main mode, was reported previously~\cite{Carlotti2019} for arrays of single-layer nanodiscs and was attributed to the edge-mode resonance. For our arrays of SAF nanodiscs, we observe two higher-field modes, both of which we associate with higher-order edge-mode resonances in the SAF nanostructure -- the interpretation supported by our micromagnetic simulations, detailed below. We note that any FMR modes related to potential domains or vortices forming in the SAF particles are excluded from consideration since the resonance fields of L$_m$, L$_{e1}$, and L$_{e2}$ ($H_r > 1$~kOe) are much higher than the SAF saturation field [$H_s \approx 300$~Oe; Fig.~\ref{fig1}(c)].

The resonance field, $H_r$, of each resonance line show a pronounced angular dependence, as seen in Fig.~\ref{fig2}(b). All lines show bi-axial anisotropy with, however, opposite hard and easy axes for L$_m$ and L$_{e1,e2}$. While L$_m$ exhibits easy axes (minimum $H_r$) at 45$\degree$ with respect to the main $x$ and $y$ axes of the array [Fig.~\ref{fig1}(a)], the edge-modes, L$_{e1}$ and L$_{e2}$, have easy axes along $x$ and $y$. Such anisotropic behavior was explained in Refs.~\onlinecite{Mathieu1997,Kakazei2015} by the inter-element dipole coupling leading to a non-uniform effective field distribution within the individual discs, even at rather high external fields. 

In order to explain the origin and behavior of the three FMR modes observed, we performed detailed micromagnetic simulations~\cite{McMichael2005}, using MuMax3~\cite{Vansteenkiste2014,Exl2014}. The layout of each element as well as the base diameter ($d_0$) used were those observed experimentally [Fig.~\ref{fig1}(c)]. Figure~\ref{fig2}(c) displays the calculated spectra for two isolated single-layer Py discs (7~nm thick) with diameters $d_0 = 150$~nm (blue line) and $d = 0.85d_0$ (red dashed line). This difference in size corresponds to the SEM-measured difference in the diameters of the bottom and top layers of our SAF nanodiscs; cf. Fig.~\ref{fig1}(b). Both simulated spectra show two resonance lines corresponding to the main and edge-resonance modes, illustrated by the insets to Fig.~\ref{fig2}(c). Of note is that despite line L$_m$ is associated with the uniform mode of a continuous Py film, the related excitation within a nanodisc is not fully spatially uniform, which is why we call it \textit{main} rather than \textit{uniform} FMR mode. Importantly, neither spectrum shows any trace of a third line observed on the experiment, nor a related additional line can be obtained when the two spectra are superposed.

The simulated spectra are substantially different when the two discs of $d_0 = 150$~nm and $d = 0.85d_0$ form a SAF element, as shown in Fig.~\ref{fig2}(d). Three resonance lines are visible, with the positions and intensities resembling the experimental data. The insets to Fig.~\ref{fig2}(d) clarify why we associate the high-field lines with edge-mode resonances. Whereas the L$_{e1}$ mode is excited mostly in the bottom layer, the L$_{e2}$ mode is more intensive in the top layer. Importantly, when the top and bottom layers are of equal diameter, only two modes, L$_m$ and L$_{e1}$, are excited; the third mode appears only when the top and bottom diameters become noticeably different (by 5--10~\%).

Figure~\ref{fig2}(e) shows the simulated angle-dependence of the resonance fields of the L$_m$, L$_{e1}$, and L$_{e2}$ modes obtained for an array of SAF nanodiscs. The simulations are in excellent agreement with the experimental resonance-field versus angle data of Fig.~\ref{fig2}(b), which further validates our interpretation of the observed modes and their properties.

\begin{figure}
\includegraphics[width=8.5 cm]{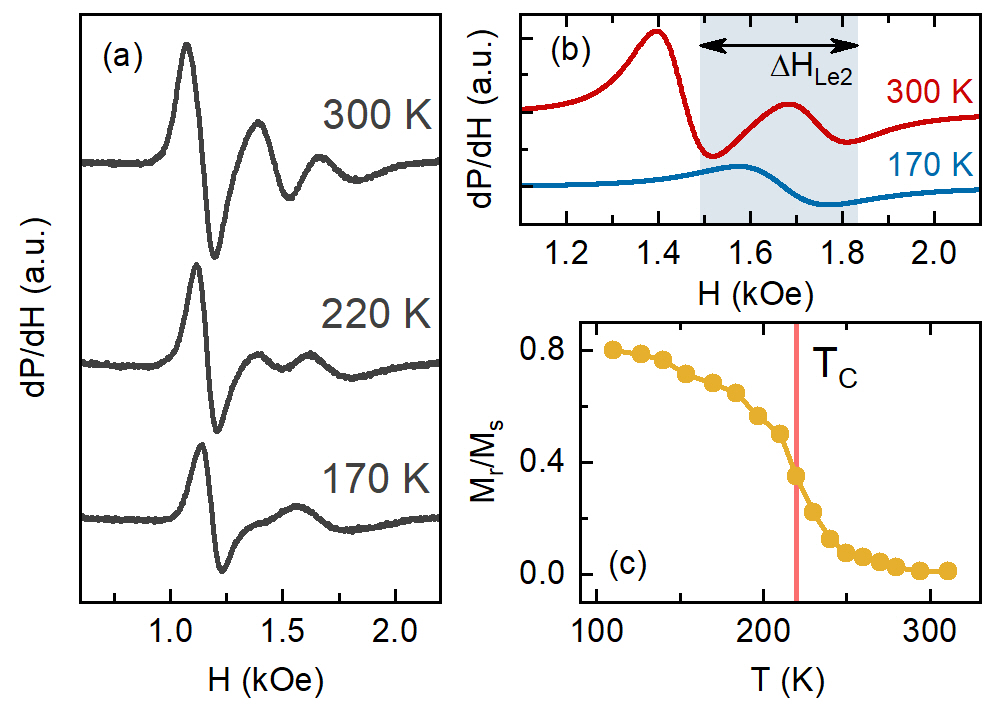}
\caption{(a) FMR spectra of SAF arrays measured at select temperatures. (b) Fits of high-order resonances (L$_{e1}$,L$_{e2}$) at above and below spacer's Curie point. (c) Normalized remnant magnetization versus temperature obtained from MOKE loops.}
\label{fig3}
\end{figure}

Figure~\ref{fig3}(a) compares the SAF FMR spectra measured as the temperature is lowered from room to below the Curie point of the NiCu spacer. Our SAF design is rather unique as it effectively transforms the nano-elements from trilayers to single-layers when the spacer transitions from para- to ferromagnetic state at about 220~K~\cite{Kravets2012,Kravets2014,Kravets2015,Kravets2016}. The associated transition in the FMR spectrum is very illuminating -- the three-mode resonance is reduced to effectively two modes, which is a known characteristic of single-layer nano-arrays. Figure~\ref{fig3}(b) shows the fits to the extracted high-order resonances (after subtracting the main peak): the pronounced double-resonance at room temperature with dominating L$_{e1}$ (red) and essentially a single broad peak at 170~K where the L$_{e1}$ peak is <1\% of its RT value. The transition from two high-order peaks to one takes place precisely at the Curie temperature of the spacer, deduced independently from our MOKE magnetization data shown in Fig.~\ref{fig3}(c). We thus confirm, in a direct experiment, that the origin of the double spin-wave resonance is unique to SAF nano-arrays. Specifically, the high-order (non-uniform) double resonance is a result of hybridization of predominantly acoustic and optical (in-phase and out-phase) oscillations in the two ferromagnetic layers comprising the SAF. These modes collapse into one on direct exchange coupling the outer layers below the $T_C$ of the spacer and, generally, can be tuned by changing the SAF geometry and the spacer properties (via magnetic dilution and/or temperature).

The above results demonstrate that the magnetization dynamics of periodic arrays of SAF nanomagnets have key distinctions from the known behavior of single-layered nano-arrays. Our variable field-angle and temperature FMR and MOKE studies, supported by morphological characterization and micromagnetic simulations, enable us to identify the origins of the distinct SAF properties observed. The three FMR modes and their pronounced in-plane anisotropy are two such distinct properties that, however, are governed by different factors. Our micromagnetic simulations indicate that the splitting of the higher-order edge mode into two is due to the dipolar coupling within the individual, asymmetric SAF particles. Here, the inter-particle interaction within the array is a secondary effect, which on the other hand dominates the anisotropic properties of the array. We find that the high-order spin excitations are hybridized acoustic (predominantly in-phase for L$_{e1}$) and optical (predominantly out-of-phase for L$_{e2}$) oscillations in the two layers comprising the SAF.

The occurrence of similar edge modes in single-layer nanodiscs is usually explained by a non-uniform internal field distribution, with its minima located at the edges orthogonal to the static magnetization of the discs~\cite{Pauselli2017}. Following similar logic for our asymmetric SAFs, we associate the two edge resonances with non-uniform field distributions that are different in each of the two nanodiscs owing to their different internal and external (inter-layer) demagnetization field profiles. The demagnetization is stronger in the smaller-diameter top layer, which results in a higher external field needed to excite the edge mode in this layer. In a symmetric SAF, with symmetric demagnetization fields, only one edge mode is excited.

The four-fold anisotropy of the three FMR modes is due to the bi-axial symmetry of our periodic arrays forming a square lattice. For our circular 150-nm SAF nanodisc arrays, this anisotropy is clearly caused by the inter-particle dipolar interactions. The observed 45$\degree$-angle between the easy axes of the main mode and the edge modes can be explained by the difference in the spacing of the respective excitation nodes for single-layer nanodot arrays~\cite{Kakazei2015}. We are able to recreate this type of anisotropy in our micromagnetic simulations of SAF nano-arrays (to be discussed elsewhere).

In summary, we have investigated experimentally and micromagnetically the spin dynamics of arrays of three-layer SAF nanoparticles where the intra-SAF coupling can be in-situ controlled by varying temperature. The results show how the intra- and inter-particle dipolar interactions combine to produce rather unique GHz properties of the system, such as additional spin-wave modes that can be tuned by the SAF geometry and/or temperature, which are relevant for the emerging field of nanostructured magnetic metamaterials and magnonic devices.

\begin{acknowledgments}
We thank Dr. Roman Verba for fruitful discussions. Support from the Swedish Research Council (VR 2018-03526), the Olle Engkvist Foundation (2020-207-0460), the Volkswagen Foundation (90418), and the National Academy of Sciences of Ukraine (Projects 0119U100469, and 0120U100457) are gratefully acknowledged.
\end{acknowledgments}

\section*{\label{sec:data}Data Availability}
The data supporting the findings of this study are available from the corresponding author upon request.

\bibliography{AoSAFs}

\begin{thebibliography}{39}%
\makeatletter
\providecommand \@ifxundefined [1]{%
 \@ifx{#1\undefined}
}%
\providecommand \@ifnum [1]{%
 \ifnum #1\expandafter \@firstoftwo
 \else \expandafter \@secondoftwo
 \fi
}%
\providecommand \@ifx [1]{%
 \ifx #1\expandafter \@firstoftwo
 \else \expandafter \@secondoftwo
 \fi
}%
\providecommand \natexlab [1]{#1}%
\providecommand \enquote  [1]{``#1''}%
\providecommand \bibnamefont  [1]{#1}%
\providecommand \bibfnamefont [1]{#1}%
\providecommand \citenamefont [1]{#1}%
\providecommand \href@noop [0]{\@secondoftwo}%
\providecommand \href [0]{\begingroup \@sanitize@url \@href}%
\providecommand \@href[1]{\@@startlink{#1}\@@href}%
\providecommand \@@href[1]{\endgroup#1\@@endlink}%
\providecommand \@sanitize@url [0]{\catcode `\\12\catcode `\$12\catcode
  `\&12\catcode `\#12\catcode `\^12\catcode `\_12\catcode `\%12\relax}%
\providecommand \@@startlink[1]{}%
\providecommand \@@endlink[0]{}%
\providecommand \url  [0]{\begingroup\@sanitize@url \@url }%
\providecommand \@url [1]{\endgroup\@href {#1}{\urlprefix }}%
\providecommand \urlprefix  [0]{URL }%
\providecommand \Eprint [0]{\href }%
\providecommand \doibase [0]{http://dx.doi.org/}%
\providecommand \selectlanguage [0]{\@gobble}%
\providecommand \bibinfo  [0]{\@secondoftwo}%
\providecommand \bibfield  [0]{\@secondoftwo}%
\providecommand \translation [1]{[#1]}%
\providecommand \BibitemOpen [0]{}%
\providecommand \bibitemStop [0]{}%
\providecommand \bibitemNoStop [0]{.\EOS\space}%
\providecommand \EOS [0]{\spacefactor3000\relax}%
\providecommand \BibitemShut  [1]{\csname bibitem#1\endcsname}%
\let\auto@bib@innerbib\@empty
\bibitem [{\citenamefont {Chumak}\ \emph {et~al.}(2015)\citenamefont {Chumak},
  \citenamefont {Vasyuchka}, \citenamefont {Serga},\ and\ \citenamefont
  {Hillebrands}}]{Chumak2015}%
  \BibitemOpen
  \bibfield  {author} {\bibinfo {author} {\bibfnamefont {A.~V.}\ \bibnamefont
  {Chumak}}, \bibinfo {author} {\bibfnamefont {V.~I.}\ \bibnamefont
  {Vasyuchka}}, \bibinfo {author} {\bibfnamefont {A.~A.}\ \bibnamefont
  {Serga}}, \ and\ \bibinfo {author} {\bibfnamefont {B.}~\bibnamefont
  {Hillebrands}},\ }\bibfield  {title} {\enquote {\bibinfo {title} {Magnon
  spintronics},}\ }\href {\doibase 10.1038/nphys3347} {\bibfield  {journal}
  {\bibinfo  {journal} {Nature Physics}\ }\textbf {\bibinfo {volume} {11}},\
  \bibinfo {pages} {453--461} (\bibinfo {year} {2015})}\BibitemShut {NoStop}%
\bibitem [{\citenamefont {Liu}\ \emph {et~al.}(2013)\citenamefont {Liu},
  \citenamefont {Ding}, \citenamefont {Kakazei},\ and\ \citenamefont
  {Adeyeye}}]{Liu2013}%
  \BibitemOpen
  \bibfield  {author} {\bibinfo {author} {\bibfnamefont {X.~M.}\ \bibnamefont
  {Liu}}, \bibinfo {author} {\bibfnamefont {J.}~\bibnamefont {Ding}}, \bibinfo
  {author} {\bibfnamefont {G.~N.}\ \bibnamefont {Kakazei}}, \ and\ \bibinfo
  {author} {\bibfnamefont {A.~O.}\ \bibnamefont {Adeyeye}},\ }\bibfield
  {title} {\enquote {\bibinfo {title} {Magnonic crystals composed of {Ni80Fe20}
  film on top of {Ni80Fe20} two-dimensional dot array},}\ }\href {\doibase
  10.1063/1.4817798} {\bibfield  {journal} {\bibinfo  {journal} {Applied
  Physics Letters}\ }\textbf {\bibinfo {volume} {103}},\ \bibinfo {pages}
  {062401} (\bibinfo {year} {2013})}\BibitemShut {NoStop}%
\bibitem [{\citenamefont {Lenk}\ \emph {et~al.}(2011)\citenamefont {Lenk},
  \citenamefont {Ulrichs}, \citenamefont {Garbs},\ and\ \citenamefont
  {M{\"{u}}nzenberg}}]{Lenk2011}%
  \BibitemOpen
  \bibfield  {author} {\bibinfo {author} {\bibfnamefont {B.}~\bibnamefont
  {Lenk}}, \bibinfo {author} {\bibfnamefont {H.}~\bibnamefont {Ulrichs}},
  \bibinfo {author} {\bibfnamefont {F.}~\bibnamefont {Garbs}}, \ and\ \bibinfo
  {author} {\bibfnamefont {M.}~\bibnamefont {M{\"{u}}nzenberg}},\ }\bibfield
  {title} {\enquote {\bibinfo {title} {The building blocks of magnonics},}\
  }\href {\doibase 10.1016/j.physrep.2011.06.003} {\bibfield  {journal}
  {\bibinfo  {journal} {Physics Reports}\ }\textbf {\bibinfo {volume} {507}},\
  \bibinfo {pages} {107--136} (\bibinfo {year} {2011})}\BibitemShut {NoStop}%
\bibitem [{\citenamefont {Haldar}\ and\ \citenamefont
  {Adeyeye}(2016)}]{Haldar2016}%
  \BibitemOpen
  \bibfield  {author} {\bibinfo {author} {\bibfnamefont {Arabinda}\
  \bibnamefont {Haldar}}\ and\ \bibinfo {author} {\bibfnamefont
  {Adekunle~Olusola}\ \bibnamefont {Adeyeye}},\ }\bibfield  {title} {\enquote
  {\bibinfo {title} {Deterministic control of magnetization dynamics in
  reconfigurable nanomagnetic networks for logic applications},}\ }\href
  {\doibase 10.1021/acsnano.5b07849} {\bibfield  {journal} {\bibinfo  {journal}
  {{ACS} Nano}\ }\textbf {\bibinfo {volume} {10}},\ \bibinfo {pages}
  {1690--1698} (\bibinfo {year} {2016})}\BibitemShut {NoStop}%
\bibitem [{\citenamefont {Saha}\ \emph {et~al.}(2012)\citenamefont {Saha},
  \citenamefont {Mandal}, \citenamefont {Barman}, \citenamefont {Kumar},
  \citenamefont {Rana}, \citenamefont {Fukuma}, \citenamefont {Sugimoto},
  \citenamefont {Otani},\ and\ \citenamefont {Barman}}]{Saha2012}%
  \BibitemOpen
  \bibfield  {author} {\bibinfo {author} {\bibfnamefont {Susmita}\ \bibnamefont
  {Saha}}, \bibinfo {author} {\bibfnamefont {Ruma}\ \bibnamefont {Mandal}},
  \bibinfo {author} {\bibfnamefont {Saswati}\ \bibnamefont {Barman}}, \bibinfo
  {author} {\bibfnamefont {Dheeraj}\ \bibnamefont {Kumar}}, \bibinfo {author}
  {\bibfnamefont {Bivas}\ \bibnamefont {Rana}}, \bibinfo {author}
  {\bibfnamefont {Yasuhiro}\ \bibnamefont {Fukuma}}, \bibinfo {author}
  {\bibfnamefont {Satoshi}\ \bibnamefont {Sugimoto}}, \bibinfo {author}
  {\bibfnamefont {YoshiChika}\ \bibnamefont {Otani}}, \ and\ \bibinfo {author}
  {\bibfnamefont {Anjan}\ \bibnamefont {Barman}},\ }\bibfield  {title}
  {\enquote {\bibinfo {title} {Tunable magnonic spectra in two-dimensional
  magnonic crystals with variable lattice symmetry},}\ }\href {\doibase
  10.1002/adfm.201202545} {\bibfield  {journal} {\bibinfo  {journal} {Advanced
  Functional Materials}\ }\textbf {\bibinfo {volume} {23}},\ \bibinfo {pages}
  {2378--2386} (\bibinfo {year} {2012})}\BibitemShut {NoStop}%
\bibitem [{\citenamefont {Liu}\ \emph {et~al.}(2008)\citenamefont {Liu},
  \citenamefont {Sydora},\ and\ \citenamefont {Freeman}}]{Liu2008}%
  \BibitemOpen
  \bibfield  {author} {\bibinfo {author} {\bibfnamefont {Zhigang}\ \bibnamefont
  {Liu}}, \bibinfo {author} {\bibfnamefont {Richard~D.}\ \bibnamefont
  {Sydora}}, \ and\ \bibinfo {author} {\bibfnamefont {Mark~R.}\ \bibnamefont
  {Freeman}},\ }\bibfield  {title} {\enquote {\bibinfo {title} {Shape effects
  on magnetization state transitions in individual 160-nm diameter {Permalloy}
  disks},}\ }\href {\doibase 10.1103/physrevb.77.174410} {\bibfield  {journal}
  {\bibinfo  {journal} {Physical Review B}\ }\textbf {\bibinfo {volume} {77}},\
  \bibinfo {pages} {174410} (\bibinfo {year} {2008})}\BibitemShut {NoStop}%
\bibitem [{\citenamefont {Schneider}\ \emph {et~al.}(2007)\citenamefont
  {Schneider}, \citenamefont {Shaw}, \citenamefont {Kos}, \citenamefont
  {Gerrits}, \citenamefont {Silva},\ and\ \citenamefont
  {McMichael}}]{Schneider2007}%
  \BibitemOpen
  \bibfield  {author} {\bibinfo {author} {\bibfnamefont {M.~L.}\ \bibnamefont
  {Schneider}}, \bibinfo {author} {\bibfnamefont {J.~M.}\ \bibnamefont {Shaw}},
  \bibinfo {author} {\bibfnamefont {A.~B.}\ \bibnamefont {Kos}}, \bibinfo
  {author} {\bibfnamefont {Th.}\ \bibnamefont {Gerrits}}, \bibinfo {author}
  {\bibfnamefont {T.~J.}\ \bibnamefont {Silva}}, \ and\ \bibinfo {author}
  {\bibfnamefont {R.~D.}\ \bibnamefont {McMichael}},\ }\bibfield  {title}
  {\enquote {\bibinfo {title} {Spin dynamics and damping in nanomagnets
  measured directly by frequency-resolved magneto-optic {Kerr} effect},}\
  }\href {\doibase 10.1063/1.2812541} {\bibfield  {journal} {\bibinfo
  {journal} {Journal of Applied Physics}\ }\textbf {\bibinfo {volume} {102}},\
  \bibinfo {pages} {103909} (\bibinfo {year} {2007})}\BibitemShut {NoStop}%
\bibitem [{\citenamefont {Verba}\ \emph {et~al.}(2020)\citenamefont {Verba},
  \citenamefont {Galkina}, \citenamefont {Tiberkevich}, \citenamefont
  {Slavin},\ and\ \citenamefont {Ivanov}}]{Verba2020}%
  \BibitemOpen
  \bibfield  {author} {\bibinfo {author} {\bibfnamefont {R.~V.}\ \bibnamefont
  {Verba}}, \bibinfo {author} {\bibfnamefont {E.~G.}\ \bibnamefont {Galkina}},
  \bibinfo {author} {\bibfnamefont {V.~S.}\ \bibnamefont {Tiberkevich}},
  \bibinfo {author} {\bibfnamefont {A.~N.}\ \bibnamefont {Slavin}}, \ and\
  \bibinfo {author} {\bibfnamefont {B.~A.}\ \bibnamefont {Ivanov}},\ }\bibfield
   {title} {\enquote {\bibinfo {title} {Spin-wave modes localized on isolated
  defects in a two-dimensional array of dipolarly coupled magnetic nanodots},}\
  }\href {\doibase 10.1103/physrevb.102.054421} {\bibfield  {journal} {\bibinfo
   {journal} {Physical Review B}\ }\textbf {\bibinfo {volume} {102}},\ \bibinfo
  {pages} {054421} (\bibinfo {year} {2020})}\BibitemShut {NoStop}%
\bibitem [{\citenamefont {Mathieu}\ \emph {et~al.}(1997)\citenamefont
  {Mathieu}, \citenamefont {Hartmann}, \citenamefont {Bauer}, \citenamefont
  {Buettner}, \citenamefont {Riedling}, \citenamefont {Roos}, \citenamefont
  {Demokritov}, \citenamefont {Hillebrands}, \citenamefont {Bartenlian},
  \citenamefont {Chappert}, \citenamefont {Decanini}, \citenamefont
  {Rousseaux}, \citenamefont {Cambril}, \citenamefont {M{\"{u}}ller},
  \citenamefont {Hoffmann},\ and\ \citenamefont {Hartmann}}]{Mathieu1997}%
  \BibitemOpen
  \bibfield  {author} {\bibinfo {author} {\bibfnamefont {C.}~\bibnamefont
  {Mathieu}}, \bibinfo {author} {\bibfnamefont {C.}~\bibnamefont {Hartmann}},
  \bibinfo {author} {\bibfnamefont {M.}~\bibnamefont {Bauer}}, \bibinfo
  {author} {\bibfnamefont {O.}~\bibnamefont {Buettner}}, \bibinfo {author}
  {\bibfnamefont {S.}~\bibnamefont {Riedling}}, \bibinfo {author}
  {\bibfnamefont {B.}~\bibnamefont {Roos}}, \bibinfo {author} {\bibfnamefont
  {S.~O.}\ \bibnamefont {Demokritov}}, \bibinfo {author} {\bibfnamefont
  {B.}~\bibnamefont {Hillebrands}}, \bibinfo {author} {\bibfnamefont
  {B.}~\bibnamefont {Bartenlian}}, \bibinfo {author} {\bibfnamefont
  {C.}~\bibnamefont {Chappert}}, \bibinfo {author} {\bibfnamefont
  {D.}~\bibnamefont {Decanini}}, \bibinfo {author} {\bibfnamefont
  {F.}~\bibnamefont {Rousseaux}}, \bibinfo {author} {\bibfnamefont
  {E.}~\bibnamefont {Cambril}}, \bibinfo {author} {\bibfnamefont
  {A.}~\bibnamefont {M{\"{u}}ller}}, \bibinfo {author} {\bibfnamefont
  {B.}~\bibnamefont {Hoffmann}}, \ and\ \bibinfo {author} {\bibfnamefont
  {U.}~\bibnamefont {Hartmann}},\ }\bibfield  {title} {\enquote {\bibinfo
  {title} {Anisotropic magnetic coupling of permalloy micron dots forming a
  square lattice},}\ }\href {\doibase 10.1063/1.119051} {\bibfield  {journal}
  {\bibinfo  {journal} {Applied Physics Letters}\ }\textbf {\bibinfo {volume}
  {70}},\ \bibinfo {pages} {2912--2914} (\bibinfo {year} {1997})}\BibitemShut
  {NoStop}%
\bibitem [{\citenamefont {Kakazei}\ \emph {et~al.}(2006)\citenamefont
  {Kakazei}, \citenamefont {Pogorelov}, \citenamefont {Costa}, \citenamefont
  {Mewes}, \citenamefont {Wigen}, \citenamefont {Hammel}, \citenamefont
  {Golub}, \citenamefont {Okuno},\ and\ \citenamefont {Novosad}}]{Kakazei2006}%
  \BibitemOpen
  \bibfield  {author} {\bibinfo {author} {\bibfnamefont {G.~N.}\ \bibnamefont
  {Kakazei}}, \bibinfo {author} {\bibfnamefont {Yu.~G.}\ \bibnamefont
  {Pogorelov}}, \bibinfo {author} {\bibfnamefont {M.~D.}\ \bibnamefont
  {Costa}}, \bibinfo {author} {\bibfnamefont {T.}~\bibnamefont {Mewes}},
  \bibinfo {author} {\bibfnamefont {P.~E.}\ \bibnamefont {Wigen}}, \bibinfo
  {author} {\bibfnamefont {P.~C.}\ \bibnamefont {Hammel}}, \bibinfo {author}
  {\bibfnamefont {V.~O.}\ \bibnamefont {Golub}}, \bibinfo {author}
  {\bibfnamefont {T.}~\bibnamefont {Okuno}}, \ and\ \bibinfo {author}
  {\bibfnamefont {V.}~\bibnamefont {Novosad}},\ }\bibfield  {title} {\enquote
  {\bibinfo {title} {Origin of fourfold anisotropy in square lattices of
  circular ferromagnetic dots},}\ }\href {\doibase 10.1103/physrevb.74.060406}
  {\bibfield  {journal} {\bibinfo  {journal} {Physical Review B}\ }\textbf
  {\bibinfo {volume} {74}},\ \bibinfo {pages} {060406} (\bibinfo {year}
  {2006})}\BibitemShut {NoStop}%
\bibitem [{\citenamefont {Carlotti}(2019)}]{Carlotti2019}%
  \BibitemOpen
  \bibfield  {author} {\bibinfo {author} {\bibfnamefont {Giovanni}\
  \bibnamefont {Carlotti}},\ }\bibfield  {title} {\enquote {\bibinfo {title}
  {Pushing down the lateral dimension of single and coupled magnetic dots to
  the nanometric scale: Characteristics and evolution of the spin-wave
  eigenmodes},}\ }\href {\doibase 10.1063/1.5110434} {\bibfield  {journal}
  {\bibinfo  {journal} {Applied Physics Reviews}\ }\textbf {\bibinfo {volume}
  {6}},\ \bibinfo {pages} {031304} (\bibinfo {year} {2019})}\BibitemShut
  {NoStop}%
\bibitem [{\citenamefont {Duine}\ \emph {et~al.}(2018)\citenamefont {Duine},
  \citenamefont {Lee}, \citenamefont {Parkin},\ and\ \citenamefont
  {Stiles}}]{Duine2018}%
  \BibitemOpen
  \bibfield  {author} {\bibinfo {author} {\bibfnamefont {R.~A.}\ \bibnamefont
  {Duine}}, \bibinfo {author} {\bibfnamefont {Kyung-Jin}\ \bibnamefont {Lee}},
  \bibinfo {author} {\bibfnamefont {Stuart S.~P.}\ \bibnamefont {Parkin}}, \
  and\ \bibinfo {author} {\bibfnamefont {M.~D.}\ \bibnamefont {Stiles}},\
  }\bibfield  {title} {\enquote {\bibinfo {title} {Synthetic antiferromagnetic
  spintronics},}\ }\href {\doibase 10.1038/s41567-018-0050-y} {\bibfield
  {journal} {\bibinfo  {journal} {Nature Physics}\ }\textbf {\bibinfo {volume}
  {14}},\ \bibinfo {pages} {217--219} (\bibinfo {year} {2018})}\BibitemShut
  {NoStop}%
\bibitem [{\citenamefont {Etesamirad}\ \emph {et~al.}(2021)\citenamefont
  {Etesamirad}, \citenamefont {Rodriguez}, \citenamefont {Bocanegra},
  \citenamefont {Verba}, \citenamefont {Katine}, \citenamefont {Krivorotov},
  \citenamefont {Tyberkevych}, \citenamefont {Ivanov},\ and\ \citenamefont
  {Barsukov}}]{Etesamirad2021}%
  \BibitemOpen
  \bibfield  {author} {\bibinfo {author} {\bibfnamefont {Arezoo}\ \bibnamefont
  {Etesamirad}}, \bibinfo {author} {\bibfnamefont {Rodolfo}\ \bibnamefont
  {Rodriguez}}, \bibinfo {author} {\bibfnamefont {Joshua}\ \bibnamefont
  {Bocanegra}}, \bibinfo {author} {\bibfnamefont {Roman}\ \bibnamefont
  {Verba}}, \bibinfo {author} {\bibfnamefont {Jordan}\ \bibnamefont {Katine}},
  \bibinfo {author} {\bibfnamefont {Ilya~N.}\ \bibnamefont {Krivorotov}},
  \bibinfo {author} {\bibfnamefont {Vasyl}\ \bibnamefont {Tyberkevych}},
  \bibinfo {author} {\bibfnamefont {Boris}\ \bibnamefont {Ivanov}}, \ and\
  \bibinfo {author} {\bibfnamefont {Igor}\ \bibnamefont {Barsukov}},\
  }\bibfield  {title} {\enquote {\bibinfo {title} {Controlling magnon
  interaction by a nanoscale switch},}\ }\href {\doibase
  10.1021/acsami.1c01562} {\bibfield  {journal} {\bibinfo  {journal} {{ACS}
  Applied Materials {\&} Interfaces}\ }\textbf {\bibinfo {volume} {13}},\
  \bibinfo {pages} {20288--20295} (\bibinfo {year} {2021})}\BibitemShut
  {NoStop}%
\bibitem [{\citenamefont {Pauselli}\ \emph {et~al.}(2017)\citenamefont
  {Pauselli}, \citenamefont {Stankiewicz},\ and\ \citenamefont
  {Carlotti}}]{Pauselli2017}%
  \BibitemOpen
  \bibfield  {author} {\bibinfo {author} {\bibfnamefont {Maurizio}\
  \bibnamefont {Pauselli}}, \bibinfo {author} {\bibfnamefont {Andrzej~A.}\
  \bibnamefont {Stankiewicz}}, \ and\ \bibinfo {author} {\bibfnamefont
  {Giovanni}\ \bibnamefont {Carlotti}},\ }\bibfield  {title} {\enquote
  {\bibinfo {title} {Linear and non-linear dynamics of the free and reference
  layers in a sub-40{\hspace{0.167em}}nm magnetic tunnel junction: a
  micromagnetic study},}\ }\href {\doibase 10.1088/1361-6463/aa8dda} {\bibfield
   {journal} {\bibinfo  {journal} {Journal of Physics D: Applied Physics}\
  }\textbf {\bibinfo {volume} {50}},\ \bibinfo {pages} {455007} (\bibinfo
  {year} {2017})}\BibitemShut {NoStop}%
\bibitem [{\citenamefont {Kamimaki}\ \emph {et~al.}(2020)\citenamefont
  {Kamimaki}, \citenamefont {Iihama}, \citenamefont {Suzuki}, \citenamefont
  {Yoshinaga},\ and\ \citenamefont {Mizukami}}]{Kamimaki2020}%
  \BibitemOpen
  \bibfield  {author} {\bibinfo {author} {\bibfnamefont {A.}~\bibnamefont
  {Kamimaki}}, \bibinfo {author} {\bibfnamefont {S.}~\bibnamefont {Iihama}},
  \bibinfo {author} {\bibfnamefont {K.~Z.}\ \bibnamefont {Suzuki}}, \bibinfo
  {author} {\bibfnamefont {N.}~\bibnamefont {Yoshinaga}}, \ and\ \bibinfo
  {author} {\bibfnamefont {S.}~\bibnamefont {Mizukami}},\ }\bibfield  {title}
  {\enquote {\bibinfo {title} {Parametric amplification of magnons in synthetic
  antiferromagnets},}\ }\href {\doibase 10.1103/physrevapplied.13.044036}
  {\bibfield  {journal} {\bibinfo  {journal} {Physical Review Applied}\
  }\textbf {\bibinfo {volume} {13}},\ \bibinfo {pages} {044036} (\bibinfo
  {year} {2020})}\BibitemShut {NoStop}%
\bibitem [{\citenamefont {Chatterjee}\ \emph {et~al.}(2018)\citenamefont
  {Chatterjee}, \citenamefont {Auffret}, \citenamefont {Sousa}, \citenamefont
  {Coelho}, \citenamefont {Prejbeanu},\ and\ \citenamefont
  {Dieny}}]{Chatterjee2018}%
  \BibitemOpen
  \bibfield  {author} {\bibinfo {author} {\bibfnamefont {Jyotirmoy}\
  \bibnamefont {Chatterjee}}, \bibinfo {author} {\bibfnamefont {Stephane}\
  \bibnamefont {Auffret}}, \bibinfo {author} {\bibfnamefont {Ricardo}\
  \bibnamefont {Sousa}}, \bibinfo {author} {\bibfnamefont {Paulo}\ \bibnamefont
  {Coelho}}, \bibinfo {author} {\bibfnamefont {Ioan-Lucian}\ \bibnamefont
  {Prejbeanu}}, \ and\ \bibinfo {author} {\bibfnamefont {Bernard}\ \bibnamefont
  {Dieny}},\ }\bibfield  {title} {\enquote {\bibinfo {title} {Novel
  multifunctional {RKKY} coupling layer for ultrathin perpendicular synthetic
  antiferromagnet},}\ }\href {\doibase 10.1038/s41598-018-29913-6} {\bibfield
  {journal} {\bibinfo  {journal} {Scientific Reports}\ }\textbf {\bibinfo
  {volume} {8}},\ \bibinfo {pages} {1--9} (\bibinfo {year} {2018})}\BibitemShut
  {NoStop}%
\bibitem [{\citenamefont {Talapatra}\ and\ \citenamefont
  {Adeyeye}(2021)}]{Talapatra2021}%
  \BibitemOpen
  \bibfield  {author} {\bibinfo {author} {\bibfnamefont {A.}~\bibnamefont
  {Talapatra}}\ and\ \bibinfo {author} {\bibfnamefont {A.~O.}\ \bibnamefont
  {Adeyeye}},\ }\bibfield  {title} {\enquote {\bibinfo {title} {Coupled
  magnetic nanostructures: Engineering lattice configurations},}\ }\href
  {\doibase 10.1063/5.0045235} {\bibfield  {journal} {\bibinfo  {journal}
  {Applied Physics Letters}\ }\textbf {\bibinfo {volume} {118}},\ \bibinfo
  {pages} {172404} (\bibinfo {year} {2021})}\BibitemShut {NoStop}%
\bibitem [{\citenamefont {Carlotti}\ \emph {et~al.}(2015)\citenamefont
  {Carlotti}, \citenamefont {Tacchi}, \citenamefont {Gubbiotti}, \citenamefont
  {Madami}, \citenamefont {Dey}, \citenamefont {Csaba},\ and\ \citenamefont
  {Porod}}]{Carlotti2015}%
  \BibitemOpen
  \bibfield  {author} {\bibinfo {author} {\bibfnamefont {G.}~\bibnamefont
  {Carlotti}}, \bibinfo {author} {\bibfnamefont {S.}~\bibnamefont {Tacchi}},
  \bibinfo {author} {\bibfnamefont {G.}~\bibnamefont {Gubbiotti}}, \bibinfo
  {author} {\bibfnamefont {M.}~\bibnamefont {Madami}}, \bibinfo {author}
  {\bibfnamefont {H.}~\bibnamefont {Dey}}, \bibinfo {author} {\bibfnamefont
  {G.}~\bibnamefont {Csaba}}, \ and\ \bibinfo {author} {\bibfnamefont
  {W.}~\bibnamefont {Porod}},\ }\bibfield  {title} {\enquote {\bibinfo {title}
  {Spin wave eigenmodes in single and coupled sub-150{\hspace{0.167em}}nm
  rectangular permalloy dots},}\ }\href {\doibase 10.1063/1.4914878} {\bibfield
   {journal} {\bibinfo  {journal} {Journal of Applied Physics}\ }\textbf
  {\bibinfo {volume} {117}},\ \bibinfo {pages} {17A316} (\bibinfo {year}
  {2015})}\BibitemShut {NoStop}%
\bibitem [{\citenamefont {Haldar}\ and\ \citenamefont
  {Adeyeye}(2020)}]{Haldar2020}%
  \BibitemOpen
  \bibfield  {author} {\bibinfo {author} {\bibfnamefont {Arabinda}\
  \bibnamefont {Haldar}}\ and\ \bibinfo {author} {\bibfnamefont
  {Adekunle~Olusola}\ \bibnamefont {Adeyeye}},\ }\bibfield  {title} {\enquote
  {\bibinfo {title} {Reconfigurable and self-biased magnonic metamaterials},}\
  }\href {\doibase 10.1063/5.0033254} {\bibfield  {journal} {\bibinfo
  {journal} {Journal of Applied Physics}\ }\textbf {\bibinfo {volume} {128}},\
  \bibinfo {pages} {240902} (\bibinfo {year} {2020})}\BibitemShut {NoStop}%
\bibitem [{\citenamefont {Begari}\ and\ \citenamefont
  {Haldar}(2018)}]{Begari2018}%
  \BibitemOpen
  \bibfield  {author} {\bibinfo {author} {\bibfnamefont {Krishna}\ \bibnamefont
  {Begari}}\ and\ \bibinfo {author} {\bibfnamefont {Arabinda}\ \bibnamefont
  {Haldar}},\ }\bibfield  {title} {\enquote {\bibinfo {title} {Bias-free giant
  tunability of microwave properties in multilayer rhomboid nanomagnets},}\
  }\href {\doibase 10.1088/1361-6463/aac86a} {\bibfield  {journal} {\bibinfo
  {journal} {Journal of Physics D: Applied Physics}\ }\textbf {\bibinfo
  {volume} {51}},\ \bibinfo {pages} {275004} (\bibinfo {year}
  {2018})}\BibitemShut {NoStop}%
\bibitem [{\citenamefont {Gartside}\ \emph {et~al.}(2021)\citenamefont
  {Gartside}, \citenamefont {Vanstone}, \citenamefont {Dion}, \citenamefont
  {Stenning}, \citenamefont {Arroo}, \citenamefont {Kurebayashi},\ and\
  \citenamefont {Branford}}]{Gartside2021}%
  \BibitemOpen
  \bibfield  {author} {\bibinfo {author} {\bibfnamefont {Jack~C.}\ \bibnamefont
  {Gartside}}, \bibinfo {author} {\bibfnamefont {Alex}\ \bibnamefont
  {Vanstone}}, \bibinfo {author} {\bibfnamefont {Troy}\ \bibnamefont {Dion}},
  \bibinfo {author} {\bibfnamefont {Kilian~D.}\ \bibnamefont {Stenning}},
  \bibinfo {author} {\bibfnamefont {Daan~M.}\ \bibnamefont {Arroo}}, \bibinfo
  {author} {\bibfnamefont {Hidekazu}\ \bibnamefont {Kurebayashi}}, \ and\
  \bibinfo {author} {\bibfnamefont {Will~R.}\ \bibnamefont {Branford}},\
  }\bibfield  {title} {\enquote {\bibinfo {title} {Reconfigurable magnonic
  mode-hybridisation and spectral control in a bicomponent artificial spin
  ice},}\ }\href {\doibase 10.1038/s41467-021-22723-x} {\bibfield  {journal}
  {\bibinfo  {journal} {Nature Communications}\ }\textbf {\bibinfo {volume}
  {12}},\ \bibinfo {pages} {1--9} (\bibinfo {year} {2021})}\BibitemShut
  {NoStop}%
\bibitem [{\citenamefont {Nembach}\ \emph {et~al.}(2021)\citenamefont
  {Nembach}, \citenamefont {McMichael}, \citenamefont {Schneider},
  \citenamefont {Shaw},\ and\ \citenamefont {Silva}}]{Nembach2021}%
  \BibitemOpen
  \bibfield  {author} {\bibinfo {author} {\bibfnamefont {H.~T.}\ \bibnamefont
  {Nembach}}, \bibinfo {author} {\bibfnamefont {R.~D.}\ \bibnamefont
  {McMichael}}, \bibinfo {author} {\bibfnamefont {M.~L.}\ \bibnamefont
  {Schneider}}, \bibinfo {author} {\bibfnamefont {J.~M.}\ \bibnamefont {Shaw}},
  \ and\ \bibinfo {author} {\bibfnamefont {T.~J.}\ \bibnamefont {Silva}},\
  }\bibfield  {title} {\enquote {\bibinfo {title} {Comparison of measured and
  simulated spin-wave mode spectra of magnetic nanostructures},}\ }\href
  {\doibase 10.1063/5.0039188} {\bibfield  {journal} {\bibinfo  {journal}
  {Applied Physics Letters}\ }\textbf {\bibinfo {volume} {118}},\ \bibinfo
  {pages} {012408} (\bibinfo {year} {2021})}\BibitemShut {NoStop}%
\bibitem [{\citenamefont {Maranville}\ \emph {et~al.}(2007)\citenamefont
  {Maranville}, \citenamefont {McMichael},\ and\ \citenamefont
  {Abraham}}]{Maranville2007}%
  \BibitemOpen
  \bibfield  {author} {\bibinfo {author} {\bibfnamefont {Brian~B.}\
  \bibnamefont {Maranville}}, \bibinfo {author} {\bibfnamefont {Robert~D.}\
  \bibnamefont {McMichael}}, \ and\ \bibinfo {author} {\bibfnamefont
  {David~W.}\ \bibnamefont {Abraham}},\ }\bibfield  {title} {\enquote {\bibinfo
  {title} {Variation of thin film edge magnetic properties with patterning
  process conditions in {Ni80Fe20} stripes},}\ }\href {\doibase
  10.1063/1.2746406} {\bibfield  {journal} {\bibinfo  {journal} {Applied
  Physics Letters}\ }\textbf {\bibinfo {volume} {90}},\ \bibinfo {pages}
  {232504} (\bibinfo {year} {2007})}\BibitemShut {NoStop}%
\bibitem [{\citenamefont {Chia}\ \emph {et~al.}(2012)\citenamefont {Chia},
  \citenamefont {Guo}, \citenamefont {Belova},\ and\ \citenamefont
  {McMichael}}]{Chia2012}%
  \BibitemOpen
  \bibfield  {author} {\bibinfo {author} {\bibfnamefont {Han-Jong}\
  \bibnamefont {Chia}}, \bibinfo {author} {\bibfnamefont {Feng}\ \bibnamefont
  {Guo}}, \bibinfo {author} {\bibfnamefont {L.~M.}\ \bibnamefont {Belova}}, \
  and\ \bibinfo {author} {\bibfnamefont {R.~D.}\ \bibnamefont {McMichael}},\
  }\bibfield  {title} {\enquote {\bibinfo {title} {Two-dimensional
  spectroscopic imaging of individual ferromagnetic nanostripes},}\ }\href
  {\doibase 10.1103/physrevb.86.184406} {\bibfield  {journal} {\bibinfo
  {journal} {Physical Review B}\ }\textbf {\bibinfo {volume} {86}},\ \bibinfo
  {pages} {184406} (\bibinfo {year} {2012})}\BibitemShut {NoStop}%
\bibitem [{\citenamefont {Zhang}\ \emph {et~al.}(2019)\citenamefont {Zhang},
  \citenamefont {Vogel}, \citenamefont {Jungfleisch}, \citenamefont {Hoffmann},
  \citenamefont {Nie},\ and\ \citenamefont {Novosad}}]{Zhang2019}%
  \BibitemOpen
  \bibfield  {author} {\bibinfo {author} {\bibfnamefont {Zhizhi}\ \bibnamefont
  {Zhang}}, \bibinfo {author} {\bibfnamefont {Michael}\ \bibnamefont {Vogel}},
  \bibinfo {author} {\bibfnamefont {M.~Benjamin}\ \bibnamefont {Jungfleisch}},
  \bibinfo {author} {\bibfnamefont {Axel}\ \bibnamefont {Hoffmann}}, \bibinfo
  {author} {\bibfnamefont {Yan}\ \bibnamefont {Nie}}, \ and\ \bibinfo {author}
  {\bibfnamefont {Valentine}\ \bibnamefont {Novosad}},\ }\bibfield  {title}
  {\enquote {\bibinfo {title} {Tuning edge-localized spin waves in magnetic
  microstripes by proximate magnetic structures},}\ }\href {\doibase
  10.1103/physrevb.100.174434} {\bibfield  {journal} {\bibinfo  {journal}
  {Physical Review B}\ }\textbf {\bibinfo {volume} {100}},\ \bibinfo {pages}
  {174434} (\bibinfo {year} {2019})}\BibitemShut {NoStop}%
\bibitem [{\citenamefont {Hu}\ \emph {et~al.}(2015)\citenamefont {Hu},
  \citenamefont {Dey}, \citenamefont {Liebing}, \citenamefont {Schumacher},
  \citenamefont {Csaba}, \citenamefont {Orlov}, \citenamefont {Bernstein},\
  and\ \citenamefont {Porod}}]{Hu2015}%
  \BibitemOpen
  \bibfield  {author} {\bibinfo {author} {\bibfnamefont {X.~K.}\ \bibnamefont
  {Hu}}, \bibinfo {author} {\bibfnamefont {H.}~\bibnamefont {Dey}}, \bibinfo
  {author} {\bibfnamefont {N.}~\bibnamefont {Liebing}}, \bibinfo {author}
  {\bibfnamefont {H.~W.}\ \bibnamefont {Schumacher}}, \bibinfo {author}
  {\bibfnamefont {G.}~\bibnamefont {Csaba}}, \bibinfo {author} {\bibfnamefont
  {A.}~\bibnamefont {Orlov}}, \bibinfo {author} {\bibfnamefont {G.~H.}\
  \bibnamefont {Bernstein}}, \ and\ \bibinfo {author} {\bibfnamefont
  {W.}~\bibnamefont {Porod}},\ }\bibfield  {title} {\enquote {\bibinfo {title}
  {Coherent precession in arrays of dipolar-coupled soft magnetic nanodots},}\
  }\href {\doibase 10.1063/1.4923160} {\bibfield  {journal} {\bibinfo
  {journal} {Journal of Applied Physics}\ }\textbf {\bibinfo {volume} {117}},\
  \bibinfo {pages} {243905} (\bibinfo {year} {2015})}\BibitemShut {NoStop}%
\bibitem [{\citenamefont {Ding}\ \emph {et~al.}(2012)\citenamefont {Ding},
  \citenamefont {Kostylev},\ and\ \citenamefont {Adeyeye}}]{Ding2012}%
  \BibitemOpen
  \bibfield  {author} {\bibinfo {author} {\bibfnamefont {J.}~\bibnamefont
  {Ding}}, \bibinfo {author} {\bibfnamefont {M.}~\bibnamefont {Kostylev}}, \
  and\ \bibinfo {author} {\bibfnamefont {A.~O.}\ \bibnamefont {Adeyeye}},\
  }\bibfield  {title} {\enquote {\bibinfo {title} {Broadband ferromagnetic
  resonance spectroscopy of permalloy triangular nanorings},}\ }\href {\doibase
  10.1063/1.3682083} {\bibfield  {journal} {\bibinfo  {journal} {Applied
  Physics Letters}\ }\textbf {\bibinfo {volume} {100}},\ \bibinfo {pages}
  {062401} (\bibinfo {year} {2012})}\BibitemShut {NoStop}%
\bibitem [{\citenamefont {Shaw}\ \emph {et~al.}(2009)\citenamefont {Shaw},
  \citenamefont {Silva}, \citenamefont {Schneider},\ and\ \citenamefont
  {McMichael}}]{Shaw2009}%
  \BibitemOpen
  \bibfield  {author} {\bibinfo {author} {\bibfnamefont {Justin~M.}\
  \bibnamefont {Shaw}}, \bibinfo {author} {\bibfnamefont {T.~J.}\ \bibnamefont
  {Silva}}, \bibinfo {author} {\bibfnamefont {Michael~L.}\ \bibnamefont
  {Schneider}}, \ and\ \bibinfo {author} {\bibfnamefont {Robert~D.}\
  \bibnamefont {McMichael}},\ }\bibfield  {title} {\enquote {\bibinfo {title}
  {Spin dynamics and mode structure in nanomagnet arrays: Effects of size and
  thickness on linewidth and damping},}\ }\href {\doibase
  10.1103/physrevb.79.184404} {\bibfield  {journal} {\bibinfo  {journal}
  {Physical Review B}\ }\textbf {\bibinfo {volume} {79}},\ \bibinfo {pages}
  {184404} (\bibinfo {year} {2009})}\BibitemShut {NoStop}%
\bibitem [{\citenamefont {Kravets}\ \emph {et~al.}(2012)\citenamefont
  {Kravets}, \citenamefont {Timoshevskii}, \citenamefont {Yanchitsky},
  \citenamefont {Bergmann}, \citenamefont {Buhler}, \citenamefont {Andersson},\
  and\ \citenamefont {Korenivski}}]{Kravets2012}%
  \BibitemOpen
  \bibfield  {author} {\bibinfo {author} {\bibfnamefont {A.~F.}\ \bibnamefont
  {Kravets}}, \bibinfo {author} {\bibfnamefont {A.~N.}\ \bibnamefont
  {Timoshevskii}}, \bibinfo {author} {\bibfnamefont {B.~Z.}\ \bibnamefont
  {Yanchitsky}}, \bibinfo {author} {\bibfnamefont {M.~A.}\ \bibnamefont
  {Bergmann}}, \bibinfo {author} {\bibfnamefont {J.}~\bibnamefont {Buhler}},
  \bibinfo {author} {\bibfnamefont {S.}~\bibnamefont {Andersson}}, \ and\
  \bibinfo {author} {\bibfnamefont {V.}~\bibnamefont {Korenivski}},\ }\bibfield
   {title} {\enquote {\bibinfo {title} {Temperature-controlled interlayer
  exchange coupling in strong/weak ferromagnetic multilayers: {A}
  thermomagnetic {C}urie switch},}\ }\href {\doibase
  10.1103/physrevb.86.214413} {\bibfield  {journal} {\bibinfo  {journal}
  {Physical Review B}\ }\textbf {\bibinfo {volume} {86}},\ \bibinfo {pages}
  {214413} (\bibinfo {year} {2012})}\BibitemShut {NoStop}%
\bibitem [{\citenamefont {Kravets}\ \emph {et~al.}(2014)\citenamefont
  {Kravets}, \citenamefont {Dzhezherya}, \citenamefont {Tovstolytkin},
  \citenamefont {Kozak}, \citenamefont {Gryshchuk}, \citenamefont {Savina},
  \citenamefont {Pashchenko}, \citenamefont {Gnatchenko}, \citenamefont
  {Koop},\ and\ \citenamefont {Korenivski}}]{Kravets2014}%
  \BibitemOpen
  \bibfield  {author} {\bibinfo {author} {\bibfnamefont {A.~F.}\ \bibnamefont
  {Kravets}}, \bibinfo {author} {\bibfnamefont {Yu.~I.}\ \bibnamefont
  {Dzhezherya}}, \bibinfo {author} {\bibfnamefont {A.~I.}\ \bibnamefont
  {Tovstolytkin}}, \bibinfo {author} {\bibfnamefont {I.~M.}\ \bibnamefont
  {Kozak}}, \bibinfo {author} {\bibfnamefont {A.}~\bibnamefont {Gryshchuk}},
  \bibinfo {author} {\bibfnamefont {Yu.~O.}\ \bibnamefont {Savina}}, \bibinfo
  {author} {\bibfnamefont {V.~A.}\ \bibnamefont {Pashchenko}}, \bibinfo
  {author} {\bibfnamefont {S.~L.}\ \bibnamefont {Gnatchenko}}, \bibinfo
  {author} {\bibfnamefont {B.}~\bibnamefont {Koop}}, \ and\ \bibinfo {author}
  {\bibfnamefont {V.}~\bibnamefont {Korenivski}},\ }\bibfield  {title}
  {\enquote {\bibinfo {title} {Synthetic ferrimagnets with thermomagnetic
  switching},}\ }\href {\doibase 10.1103/physrevb.90.104427} {\bibfield
  {journal} {\bibinfo  {journal} {Physical Review B}\ }\textbf {\bibinfo
  {volume} {90}},\ \bibinfo {pages} {104427} (\bibinfo {year}
  {2014})}\BibitemShut {NoStop}%
\bibitem [{\citenamefont {Polishchuk}\ \emph {et~al.}(2018)\citenamefont
  {Polishchuk}, \citenamefont {Tykhonenko-Polishchuk}, \citenamefont
  {Borynskyi}, \citenamefont {Kravets}, \citenamefont {Tovstolytkin},\ and\
  \citenamefont {Korenivski}}]{Polishchuk2018}%
  \BibitemOpen
  \bibfield  {author} {\bibinfo {author} {\bibfnamefont {Dmytro}\ \bibnamefont
  {Polishchuk}}, \bibinfo {author} {\bibfnamefont {Yuliya}\ \bibnamefont
  {Tykhonenko-Polishchuk}}, \bibinfo {author} {\bibfnamefont {Vladyslav}\
  \bibnamefont {Borynskyi}}, \bibinfo {author} {\bibfnamefont {Anatolii}\
  \bibnamefont {Kravets}}, \bibinfo {author} {\bibfnamefont {Alexandr}\
  \bibnamefont {Tovstolytkin}}, \ and\ \bibinfo {author} {\bibfnamefont
  {Vladislav}\ \bibnamefont {Korenivski}},\ }\bibfield  {title} {\enquote
  {\bibinfo {title} {Magnetic hysteresis in nanostructures with thermally
  controlled {RKKY} coupling},}\ }\href {\doibase 10.1186/s11671-018-2669-0}
  {\bibfield  {journal} {\bibinfo  {journal} {Nanoscale Research Letters}\
  }\textbf {\bibinfo {volume} {13}},\ \bibinfo {pages} {1--7} (\bibinfo {year}
  {2018})}\BibitemShut {NoStop}%
\bibitem [{\citenamefont {Koop}\ \emph {et~al.}(2017)\citenamefont {Koop},
  \citenamefont {Descamps}, \citenamefont {Holmgren},\ and\ \citenamefont
  {Korenivski}}]{Koop2017}%
  \BibitemOpen
  \bibfield  {author} {\bibinfo {author} {\bibfnamefont {B.~C.}\ \bibnamefont
  {Koop}}, \bibinfo {author} {\bibfnamefont {T.}~\bibnamefont {Descamps}},
  \bibinfo {author} {\bibfnamefont {E.}~\bibnamefont {Holmgren}}, \ and\
  \bibinfo {author} {\bibfnamefont {V.}~\bibnamefont {Korenivski}},\ }\bibfield
   {title} {\enquote {\bibinfo {title} {Relaxation-free and inertial switching
  in synthetic antiferromagnets subject to super-resonant excitation},}\ }\href
  {\doibase 10.1109/tmag.2017.2707589} {\bibfield  {journal} {\bibinfo
  {journal} {{IEEE} Transactions on Magnetics}\ }\textbf {\bibinfo {volume}
  {53}},\ \bibinfo {pages} {1--5} (\bibinfo {year} {2017})}\BibitemShut
  {NoStop}%
\bibitem [{\citenamefont {Holmgren}\ \emph {et~al.}(2017)\citenamefont
  {Holmgren}, \citenamefont {Bondarenko}, \citenamefont {Koop}, \citenamefont
  {Ivanov},\ and\ \citenamefont {Korenivski}}]{Holmgren2017}%
  \BibitemOpen
  \bibfield  {author} {\bibinfo {author} {\bibfnamefont {Erik}\ \bibnamefont
  {Holmgren}}, \bibinfo {author} {\bibfnamefont {Artem}\ \bibnamefont
  {Bondarenko}}, \bibinfo {author} {\bibfnamefont {Bjorn}\ \bibnamefont
  {Koop}}, \bibinfo {author} {\bibfnamefont {Boris}\ \bibnamefont {Ivanov}}, \
  and\ \bibinfo {author} {\bibfnamefont {Vladislav}\ \bibnamefont
  {Korenivski}},\ }\bibfield  {title} {\enquote {\bibinfo {title}
  {Non-degeneracy and effects of pinning in strongly coupled vortex pairs},}\
  }\href {\doibase 10.1109/tmag.2017.2697204} {\bibfield  {journal} {\bibinfo
  {journal} {{IEEE} Transactions on Magnetics}\ }\textbf {\bibinfo {volume}
  {53}},\ \bibinfo {pages} {1--5} (\bibinfo {year} {2017})}\BibitemShut
  {NoStop}%
\bibitem [{\citenamefont {Kakazei}\ \emph {et~al.}(2015)\citenamefont
  {Kakazei}, \citenamefont {Liu}, \citenamefont {Ding}, \citenamefont {Golub},
  \citenamefont {Salyuk}, \citenamefont {Verba}, \citenamefont {Bunyaev},\ and\
  \citenamefont {Adeyeye}}]{Kakazei2015}%
  \BibitemOpen
  \bibfield  {author} {\bibinfo {author} {\bibfnamefont {G.~N.}\ \bibnamefont
  {Kakazei}}, \bibinfo {author} {\bibfnamefont {X.~M.}\ \bibnamefont {Liu}},
  \bibinfo {author} {\bibfnamefont {J.}~\bibnamefont {Ding}}, \bibinfo {author}
  {\bibfnamefont {V.~O.}\ \bibnamefont {Golub}}, \bibinfo {author}
  {\bibfnamefont {O.~Y.}\ \bibnamefont {Salyuk}}, \bibinfo {author}
  {\bibfnamefont {R.~V.}\ \bibnamefont {Verba}}, \bibinfo {author}
  {\bibfnamefont {S.~A.}\ \bibnamefont {Bunyaev}}, \ and\ \bibinfo {author}
  {\bibfnamefont {A.~O.}\ \bibnamefont {Adeyeye}},\ }\bibfield  {title}
  {\enquote {\bibinfo {title} {Large four-fold magnetic anisotropy in
  two-dimensional modulated {Ni80Fe20} films},}\ }\href {\doibase
  10.1063/1.4936994} {\bibfield  {journal} {\bibinfo  {journal} {Applied
  Physics Letters}\ }\textbf {\bibinfo {volume} {107}},\ \bibinfo {pages}
  {232402} (\bibinfo {year} {2015})}\BibitemShut {NoStop}%
\bibitem [{\citenamefont {McMichael}\ and\ \citenamefont
  {Stiles}(2005)}]{McMichael2005}%
  \BibitemOpen
  \bibfield  {author} {\bibinfo {author} {\bibfnamefont {R.~D.}\ \bibnamefont
  {McMichael}}\ and\ \bibinfo {author} {\bibfnamefont {M.~D.}\ \bibnamefont
  {Stiles}},\ }\bibfield  {title} {\enquote {\bibinfo {title} {Magnetic normal
  modes of nanoelements},}\ }\href {\doibase 10.1063/1.1852191} {\bibfield
  {journal} {\bibinfo  {journal} {Journal of Applied Physics}\ }\textbf
  {\bibinfo {volume} {97}},\ \bibinfo {pages} {10J901} (\bibinfo {year}
  {2005})}\BibitemShut {NoStop}%
\bibitem [{\citenamefont {Vansteenkiste}\ \emph {et~al.}(2014)\citenamefont
  {Vansteenkiste}, \citenamefont {Leliaert}, \citenamefont {Dvornik},
  \citenamefont {Helsen}, \citenamefont {Garcia-Sanchez},\ and\ \citenamefont
  {Waeyenberge}}]{Vansteenkiste2014}%
  \BibitemOpen
  \bibfield  {author} {\bibinfo {author} {\bibfnamefont {Arne}\ \bibnamefont
  {Vansteenkiste}}, \bibinfo {author} {\bibfnamefont {Jonathan}\ \bibnamefont
  {Leliaert}}, \bibinfo {author} {\bibfnamefont {Mykola}\ \bibnamefont
  {Dvornik}}, \bibinfo {author} {\bibfnamefont {Mathias}\ \bibnamefont
  {Helsen}}, \bibinfo {author} {\bibfnamefont {Felipe}\ \bibnamefont
  {Garcia-Sanchez}}, \ and\ \bibinfo {author} {\bibfnamefont {Bartel~Van}\
  \bibnamefont {Waeyenberge}},\ }\bibfield  {title} {\enquote {\bibinfo {title}
  {The design and verification of {MuMax}3},}\ }\href {\doibase
  10.1063/1.4899186} {\bibfield  {journal} {\bibinfo  {journal} {{AIP}
  Advances}\ }\textbf {\bibinfo {volume} {4}},\ \bibinfo {pages} {107133}
  (\bibinfo {year} {2014})}\BibitemShut {NoStop}%
\bibitem [{\citenamefont {Exl}\ \emph {et~al.}(2014)\citenamefont {Exl},
  \citenamefont {Bance}, \citenamefont {Reichel}, \citenamefont {Schrefl},
  \citenamefont {Stimming},\ and\ \citenamefont {Mauser}}]{Exl2014}%
  \BibitemOpen
  \bibfield  {author} {\bibinfo {author} {\bibfnamefont {Lukas}\ \bibnamefont
  {Exl}}, \bibinfo {author} {\bibfnamefont {Simon}\ \bibnamefont {Bance}},
  \bibinfo {author} {\bibfnamefont {Franz}\ \bibnamefont {Reichel}}, \bibinfo
  {author} {\bibfnamefont {Thomas}\ \bibnamefont {Schrefl}}, \bibinfo {author}
  {\bibfnamefont {Hans~Peter}\ \bibnamefont {Stimming}}, \ and\ \bibinfo
  {author} {\bibfnamefont {Norbert~J.}\ \bibnamefont {Mauser}},\ }\bibfield
  {title} {\enquote {\bibinfo {title} {{LaBonte\textquotesingle}s method
  revisited: An effective steepest descent method for micromagnetic energy
  minimization},}\ }\href {\doibase 10.1063/1.4862839} {\bibfield  {journal}
  {\bibinfo  {journal} {Journal of Applied Physics}\ }\textbf {\bibinfo
  {volume} {115}},\ \bibinfo {pages} {17D118} (\bibinfo {year}
  {2014})}\BibitemShut {NoStop}%
\bibitem [{\citenamefont {Kravets}\ \emph {et~al.}(2015)\citenamefont
  {Kravets}, \citenamefont {Tovstolytkin}, \citenamefont {Dzhezherya},
  \citenamefont {Polishchuk}, \citenamefont {Kozak},\ and\ \citenamefont
  {Korenivski}}]{Kravets2015}%
  \BibitemOpen
  \bibfield  {author} {\bibinfo {author} {\bibfnamefont {A.~F.}\ \bibnamefont
  {Kravets}}, \bibinfo {author} {\bibfnamefont {A.~I.}\ \bibnamefont
  {Tovstolytkin}}, \bibinfo {author} {\bibfnamefont {Yu~I.}\ \bibnamefont
  {Dzhezherya}}, \bibinfo {author} {\bibfnamefont {D.~M.}\ \bibnamefont
  {Polishchuk}}, \bibinfo {author} {\bibfnamefont {I.~M.}\ \bibnamefont
  {Kozak}}, \ and\ \bibinfo {author} {\bibfnamefont {V.}~\bibnamefont
  {Korenivski}},\ }\bibfield  {title} {\enquote {\bibinfo {title} {Spin
  dynamics in a {C}urie-switch},}\ }\href {\doibase
  10.1088/0953-8984/27/44/446003} {\bibfield  {journal} {\bibinfo  {journal}
  {Journal of Physics: Condensed Matter}\ }\textbf {\bibinfo {volume} {27}},\
  \bibinfo {pages} {446003} (\bibinfo {year} {2015})}\BibitemShut {NoStop}%
\bibitem [{\citenamefont {Kravets}\ \emph {et~al.}(2016)\citenamefont
  {Kravets}, \citenamefont {Polishchuk}, \citenamefont {Dzhezherya},
  \citenamefont {Tovstolytkin}, \citenamefont {Golub},\ and\ \citenamefont
  {Korenivski}}]{Kravets2016}%
  \BibitemOpen
  \bibfield  {author} {\bibinfo {author} {\bibfnamefont {A.~F.}\ \bibnamefont
  {Kravets}}, \bibinfo {author} {\bibfnamefont {D.~M.}\ \bibnamefont
  {Polishchuk}}, \bibinfo {author} {\bibfnamefont {Yu.~I.}\ \bibnamefont
  {Dzhezherya}}, \bibinfo {author} {\bibfnamefont {A.~I.}\ \bibnamefont
  {Tovstolytkin}}, \bibinfo {author} {\bibfnamefont {V.~O.}\ \bibnamefont
  {Golub}}, \ and\ \bibinfo {author} {\bibfnamefont {V.}~\bibnamefont
  {Korenivski}},\ }\bibfield  {title} {\enquote {\bibinfo {title} {Anisotropic
  magnetization relaxation in ferromagnetic multilayers with variable
  interlayer exchange coupling},}\ }\href {\doibase 10.1103/physrevb.94.064429}
  {\bibfield  {journal} {\bibinfo  {journal} {Physical Review B}\ }\textbf
  {\bibinfo {volume} {94}},\ \bibinfo {pages} {064429} (\bibinfo {year}
  {2016})}\BibitemShut {NoStop}%
\end{thebibliography}%

\end{document}